\RequirePackage{fix-cm}
\documentclass[twocolumn]{svjour3}          
\smartqed  
\usepackage{graphicx}
\usepackage{hyperref}
\usepackage{xcolor}
\usepackage{aas_macros}

 \journalname{Experimental Astronomy}

\begin{document}

\title{Determination of stellar parameters for Ariel targets: a comparison analysis between different spectroscopic methods.
}


\author{Anna Brucalassi  \and
        Maria Tsantaki  \and 
        Laura Magrini \and
        Sergio Sousa \and
        Camilla Danielski \and
        Katia Biazzo \and
        Giada Casali \and
        Mathieu Van der Swaelmen \and
        Monica Rainer \and
        Vardan Adibekyan \and
        Elisa Delgado-Mena \and 
        Nicoletta Sanna 
}

\institute{A. Brucalassi \at
              INAF-Osservatorio Astronomico Arcetri,\\ 
              Largo Enrico Fermi 5, 50125 Firenze \\
              Tel.: +39-055-2752 243\\
              Fax: +39-055-220039\\
              \email{anna.brucalassi@inaf.it}           
           \and
            M. Tsantaki, L. Magrini, G. Casali, M. Van der Swaelmen, M. Rainer, N. Sanna  \at
            INAF-Osservatorio Astronomico di Arcetri,\\ 
            Largo Enrico Fermi 5, 50125 Firenze, Italy\\
             \and 
             G. Casali \at
             Dipartimento di Fisica e Astronomia, Università degli Studi di Firenze, \\
             via G. Sansone 1, 50019 Sesto Fiorentino (Firenze), Italy \\
            \and 
            K. Biazzo \at
            INAF-Osservatorio Astronomico di Roma\\
            Via di Frascati 33, 00044 
            Monte Porzio Catone, Italy\\
            \and
            S. Sousa, V. Adibekyan, E. Delgado-Mena \at
            Instituto de Astrof\'{\i}sica e Ci\^{e}ncias do Espa\c{c}o, Universidade do Porto, CAUP, Rua das Estrelas, 4150-762 Porto, Portugal 
            \and
            C. Danielski \at 
            UCL Centre for Space Exochemistry Data\\
            Atlas Building,  Fermi Avenue, Harwell Campus l Didcot l OX11 0QR (UK)
}

\date{Received: date / Accepted: date}

\maketitle

\begin{abstract}
Ariel has been selected as the next ESA M4 science mission and it is expected to be launched in 2028.
During its 4-year mission, Ariel will observe the atmospheres of a large and diversified population of transiting exoplanets.
A key factor for the achievement of the scientific goal of Ariel is the selection strategy for the definition of the input target list.
A meaningful choice of the targets requires an accurate knowledge of the  planet hosting star properties and this is necessary to be obtained well before the launch.
In this work, we present the results of a bench-marking analysis between three different spectroscopic techniques used to determine stellar parameters for a selected number of targets belonging to the Ariel reference sample.
We aim to consolidate a method that will be used to homogeneously determine the stellar parameters of the complete Ariel reference sample.
Homogeneous, accurate and precise derivation of stellar parameters is crucial for characterising exoplanet-host stars and in turn is a key factor for the accuracy of the planet properties.

\end{abstract}

\section{Introduction}
\label{intro}
So far, more than 4000 planets (with more than 3000 transiting their stars) have been detected showing an incredible diversity in terms of masses, sizes and orbits. Moreover, thousands of Jupiter-size down to Earth-size planets are expected to be discovered in the next few years by Gaia \cite{Sozzetti2010,Perryman2014}, TESS \cite{Ricker2014}, CHEOPS \cite{Broeg2013} and the upcoming space surveys, such as  PLATO \cite{Rauer2016}, along with ground-based surveys, like WASP \cite{Pollacco2006}, NGTS \cite{Wheatley2013}, TRAPPIST \cite{Jehin2013}, HARPS \cite{Mayor2003} and ESPRESSO \cite{Pepe2010}, CARMENES\cite{Quirrenbach2016} and SPIRoU\cite{Artigau2016}.  
The recent success of ground-based, space transit and radial velocity searches has ushered exoplanet research into an era of \emph{characterisation} studies
with the goal to investigate the nature, formation, and evolutionary history of the detected objects.\\
At first, these studies have been focused on understanding the internal structure of exoplanets. From the transit light-curve, the planet radius can be measured and from spectroscopic Doppler measurements, the planet mass is obtained. From the bulk density we have the first hints of the internal structure of the exoplanet and the gas/ice/rock ratios.
However, to have a reliable estimate of the density, an accurate knowledge of both radius (with a precision of up to 5$\%$) and planetary mass (up to 10$\%$) is necessary \cite{Wagner2011,Bean2011}.
Additionally, the absolute value of planetary radius and mass relies on the precise determination of the radius and mass of the exoplanet-host star. The derivation of these last two values is in turn strongly connected to the effective temperature ($T_{\rm eff}$), surface gravity ($\log g$), and the metallicity of the star. Thus, the planetary properties are critically dependent on their stellar host properties \cite{Torres2012,Sousa2015,Adibekyan2018}.

Furthermore, transiting planets provide us one of the best ways of characterising their atmospheres.
In-transit spectroscopy as well as secondary transit studies \cite{Brogi2012,Stevenson2014,Sing2016,Fu2017,Tsiaras2018,Zhang2018,Pinhas2019}
using space observatories, as Hubble and Spitzer, and some ground-based observatories,
have yielded the detection of some important molecules present in the planetary atmospheres for a limited number of targets,
or have identified the presence of clouds, probing the thermal structure and providing some constraints on the planet properties. However, the data available is still too sparse to provide a consistent interpretation and the achieved results point out the main limitations of the existing facilities:
very narrow wavelength coverage, observations usually not simultaneous for a wider spectral range with the introduction of systematic noise, insufficient time allocated to exoplanet science, and more in general
the lack of a dedicated space-based exoplanet spectroscopy mission.  
Thus, our current knowledge of exoplanetary atmospheric and thermal characteristics is still very limited.

Ariel (\emph{Atmospheric Remote-sensing Infrared Exoplanet Large-survey}) has been selected as the next ESA-M4 science mission \cite{Tinetti2018} and it is expected to be launched in 2028.
During its 4-year mission, Ariel will observe the atmospheres of a statistically representative sample ($\sim$1000) of transiting gaseous (Jupiters, Saturns, Neptunes) and rocky (super-Earths and sub-Neptunes) planets 
using transit spectroscopy in the 1.10-7.8$\mu$m spectral range and three narrow-bands photometry in the optical.
The wavelength range proposed covers all the expected major atmospheric gases from H$_{2}$O, CO$_{2}$, CH$_{4}$, NH$_{3}$, HCN, H$_{2}$S up to the more exotic metallic compounds, such as TiO, VO, and condensed species.\\
Ariel is designed as a dedicated survey mission for transit, eclipse and phase-curved spectroscopy, 
 providing a homogeneous dataset, with a consistent pipeline and an well-defined target selection strategy, maximising the scientific yield.
 Focusing on transit, eclipse spectroscopy, the methods are based on the differential analysis of the star and planet spectra 
 in and out of transit, allowing to measure planetary atmospheric signals of 10-100 ppm relative to the star.
Such small signals require an exact knowledge of the host star spectrum, at least at the same level of the planetary signal, to map any stellar intrinsic variation (i.e. due to magnetic activity and convective turbulence) in order to avoid misleading results with planetary features \cite{Robertson2015}.
 
Another important point is that information on the host star composition is critical to separate the signatures left on the planet by its formation, evolution and migration processes, from those due to the specific chemistry of the host star \cite{Turrini2018}.
Indeed, recent studies suggest that planetary O/H, C/H, C/O ratios and metallicity with respect to the stellar values 
could provide stronger constraints on the planet formation region and their migration mechanisms 
(see \cite{Madhusudhan2016} and references therein for a recent review),
but similar considerations apply to other elements (e.g., N, S, Ti, Al, \cite{Turrini2018}).

Finally, an increasing number of studies have pointed towards the existence of correlations between 
the properties of the host stars and the characteristics and frequency of their planetary systems.
In this respect, the correlation between the stellar metallicity and the frequency of giant planets \cite{Santos2004,Valenti2005,Sousa2008},
the connection between radius vs metallicity \cite{Buchhave2014,Schlaufman2015},  eccentricity vs metallicity \cite{Adibekyan2013,Wilson2018},
the role of the abundances of other elements \cite{Adibekyan2012,Delgado2014,Delgado2017} in the host stars are only few examples of different results that take a clear shape as the new planet discoveries increase, shading light on many details still missing concerning planet formation and evolution.\\
Such works rely upon homogeneously and precisely derived stellar parameters.
Therefore, homogeneous derivation of stellar parameters using high-quality data is crucial for characterising
exoplanet-host stars \cite{Santos2013,Andreasen2017,Sousa2018}, and in turn, is fundamental to improve the accuracy of the planet properties.\\
It is important to underline how a well-defined target selection strategy and the definition of the input target list with a statistically significant dataset in the range of relevant stellar/planet parameters have a fundamental role for maximising the scientific yield and the achievement of scientific goals of Ariel. A meaningful choice of the targets requires an accurate study of the stellar properties that need to be derived in advance and continuously updated as the mission approaches launch and the target list evolves with the new exoplanet discoveries.\\
In this context, we have started a benchmarking analysis between three different spectroscopic techniques used 
to determine stellar parameters for selected targets belonging to the Ariel Reference Sample \cite{Edwards2019}.
Our goal is to consolidate a method that will be applied to homogeneously determine the stellar parameters for the complete Ariel Reference Sample.
More generally, we refer also to the works of \cite{Lebzelter2012} and \cite{Smiljanic2014} where comparisons between the results from different spectroscopic techniques are discussed.

For a global approach to characterize the stars of Ariel Reference Sample see also \cite{Danielski2020}, where an overview on the methods used to determine stellar fundamental parameters, elemental abundances, activity indices, and stellar ages for the Ariel Reference Sample is given and in particular, results for the homogeneous estimation of elemental abundances of Al, Mg, Si, C, N, and the activity indices  S  and log(R' HK) are presented.

In the following sections we describe the star sample analysed, we give an overview of the methodology applied to derive the stellar parameters and then we present a comparative analysis of the results.

\section{Star Sample}
\label{sec:2}
On the bases of Ariel capabilities, a list of targets (Ariel Reference Sample) to be observed during the primary mission life
was prepared for the Phase A \cite{Zingales2018,Edwards2019}. The sample includes $\sim$1000 potential targets with stellar types FGKM (typically brighter than K\,=\,11 mag) and planetary parameters in a range of size between Jupiter down to Earth-like planets, temperature in the range between 500K-2500K and bulk density between 0.10-10.0g/cm$^{3}$.

We started our analysis cross-matching the Ariel Reference Sample with the stars available in the SWEET-Cat
(Stars With ExoplanETs Catalogue \footnote{\url{https://www.astro.up.pt/resources/sweet-cat/}})
 and we selected all the common targets with parameters source "flag=1", which
are the stars analysed homogeneously (see \cite{Santos2013}).
\begin{figure}
  \includegraphics[width=0.50\textwidth]{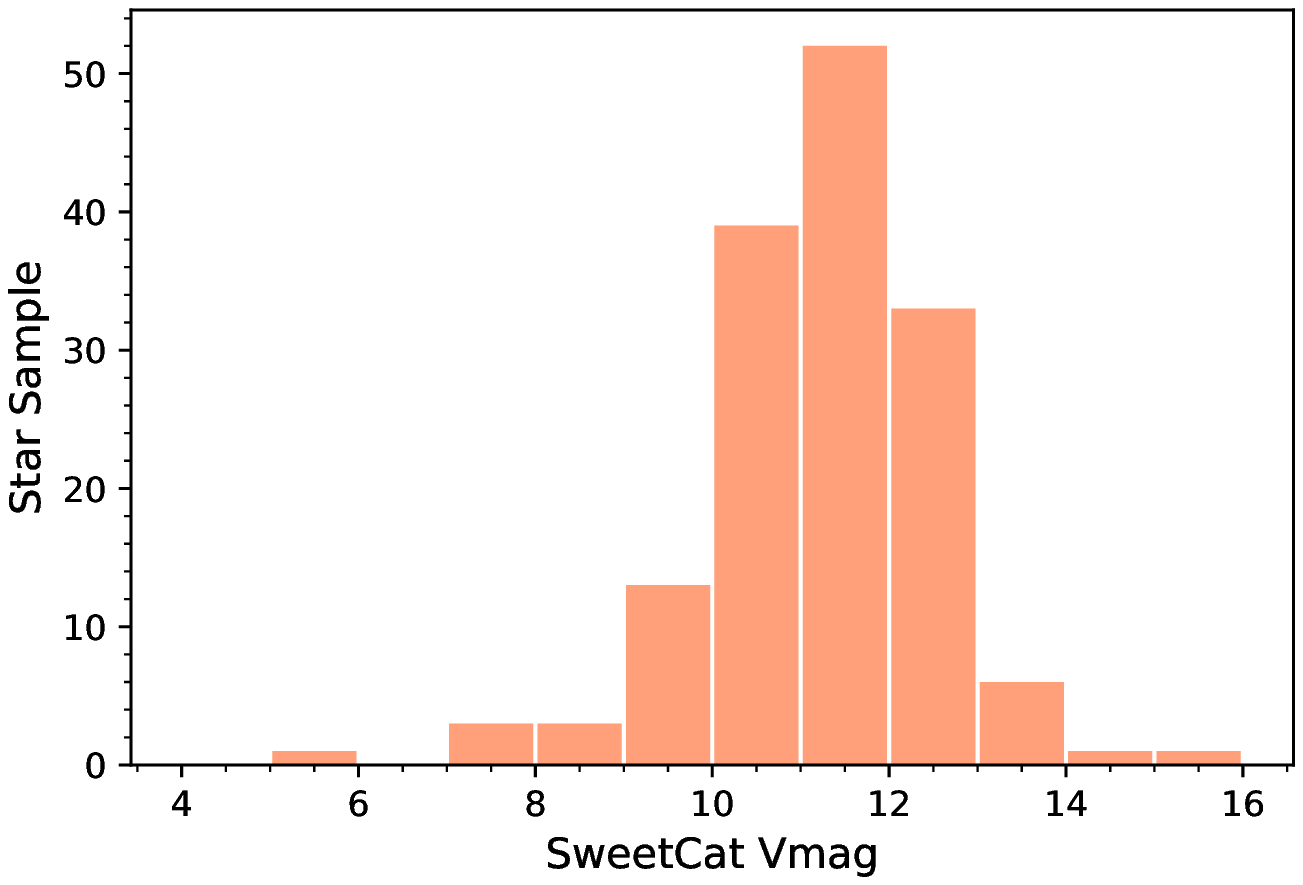}
  \includegraphics[width=0.50\textwidth]{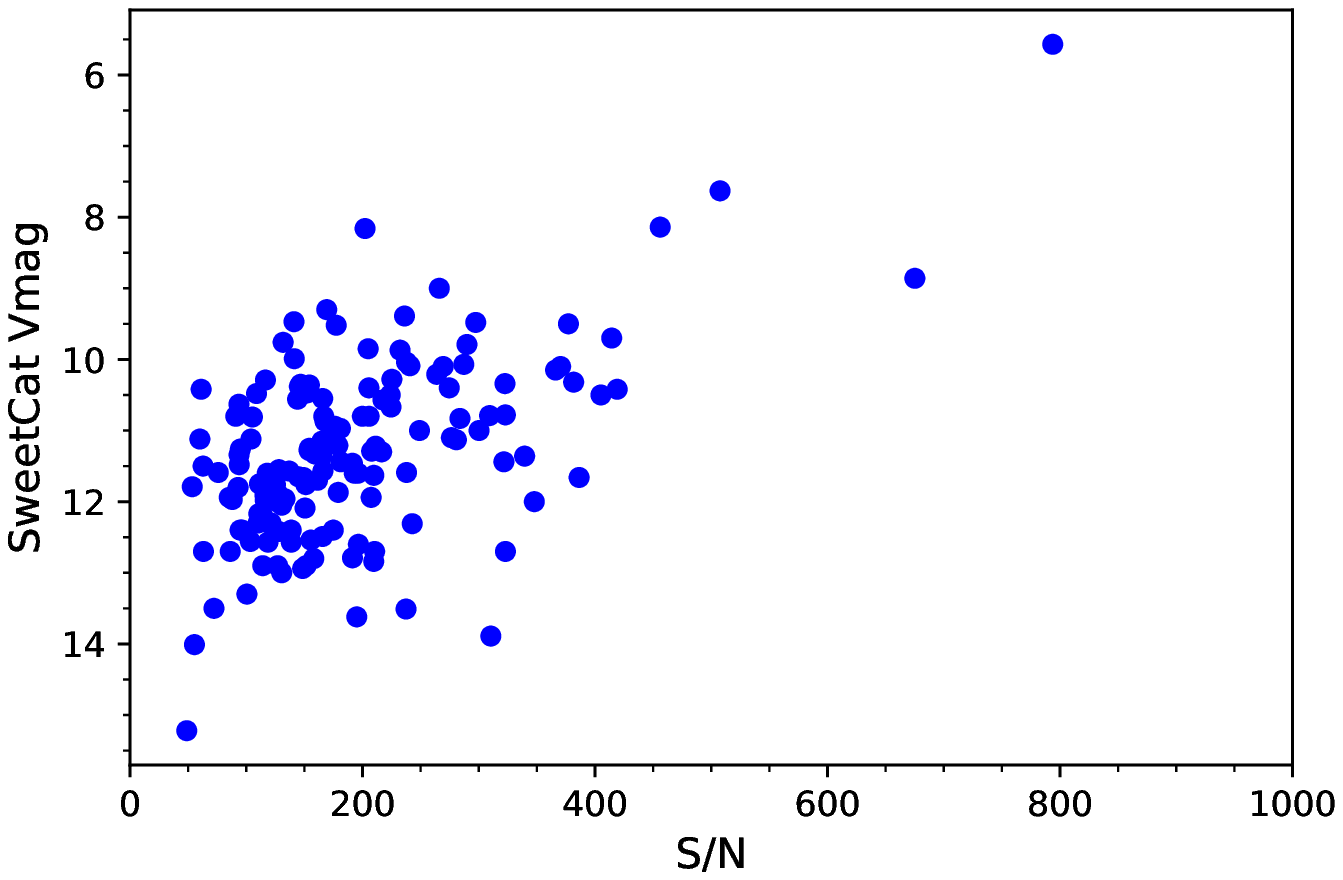}  
\caption{Top panel: Distribution of the star sample analyzed as a function of V mag listed in the SWEET-Cat catalog. Bottom panel: V mag vs S/N for our star sample.}
\label{fig:1}       
\end{figure}
Thus, our starting sub-sample includes 155 FGK stars in a range of 5$<$V(mag)$<$16
with relative spectra having a signal-to-noise ratio (S/N) between $\sim$50 and $\sim$800 (see Fig.~\ref{fig:1}).
Stellar parameters for these 155 stars available in the SWEET-Cat, were considered our baseline for the comparison analysis.
An accurate description of the spectroscopic data used to derive SWEET-Cat stellar parameters is presented in \cite{Sousa2018} and \cite{Santos2013}.

We then re-determined the stellar parameters of this sub-sample using two
different methods: FAMA (\emph{Fast Automatic MOOG Analysis}, \cite{Magrini2013}), based on the equivalent widths analysis
and FASMA (\emph{Fast Analysis of Spectra Made Automatically}, \cite{Tsantaki2018}), based on the spectral synthesis method. 
For our work, we used the same archival spectra employed for the SWEET-Cat analysis.
The characteristics of each spectrograph and the number of stars considered in our analysis for each instrument are listed in Table~\ref{tab:1}.
\begin{table}
\caption{Spectrograph information: resolving power, spectral ranges and number of stars considered.}
\label{tab:1}       
\begin{tabular}{llll}
\hline
\hline
Instrument & Resolving Power  & Spectral Range  & N \\
		 & ($\lambda / \Delta \lambda$) & ($\AA$)	&of Stars \\
\hline
ESPADONS & 80000 & 3700-10500 & 2 \\
FEROS & 48000 & 3600-9200 & 38 \\
FIES & 67000 & 3700-7300 & 7 \\
HARPS & 100000 & 3800-7000 & 39 \\
HARPSN & 115000 & 3830-6900 & 4 \\
NARVAL & 80000 & 3700-10500 & 2 \\
SARG & 57000-86000 & 5100-10100 & 5 \\
SOPHIE & 75000 & 3820-6920 & 15 \\
UVES & 110000 & 3000-6800 & 43 \\
\hline
\hline
\end{tabular}
\end{table}
Excluding stars for which FAMA and FASMA codes did not converge (spectra dominated by fringing, very low S/N, fast rotator targets, very cool stars), we obtained results for about 93\% of the sample.\\
Individual stellar parameters derived by FAMA and FASMA for each analysed star are provided as online material in the format listed in Table~\ref{tab:Summary}.

\section{Derivation of Stellar Parameters}

\subsection{SWEET-Cat}
\label{sec:2.1}
SWEET-Cat is a catalog of stellar parameters taken 
in general from the literature for planet hosting stars listed within the Extrasolar Planets Encyclopaedia \footnote{\url{http://exoplanets.eu/}}. 
This catalogue is continuously updated when new planets are announced and new stellar parameters derived. 
In particular, for stars with spectra acquired with high resolution spectrographs (mainly HARPS, UVES, FEROS) and high signal-to-noise ratios (mostly S/N$>$100), the stellar parameters are obtained in a homogeneous way (flagging the targets with 1, as mentioned above). The method used to derive stellar parameters is described in, e.g., \cite{Santos2004} and \cite{Sousa2008}.
Briefly, local thermodynamic equilibrium (LTE) condition is assumed, a grid of Kurucz plane-parallel model
atmospheres (\cite{Kurucz1993}, ATLAS9), and the ARES code \cite{Sousa2007} for measuring line equivalent widths (EWs) are considered. Stellar parameters (effective temperature T$_{\rm eff}$, surface gravity log~g, microturbulence velocity V$_{\rm micro}$, and iron abundance [Fe/H]) are derived using the MOOG code (version 2002; \cite{Sneden1973}) and the line list by \cite{Sousa2008}. Excitation and ionization equilibria of the Fe I and Fe II weak lines were imposed to derive stellar parameters. The errors on the atmospheric parameters were derived as in \cite{Santos2004} and references therein.

The stellar masses listed in the SWEET-Cat catalog were derived with the  calibration  presented  in \cite{Torres2010}, using as input the spectroscopic parameters. According to \cite{Santos2013}, a correction was applied for the cases in which the calibration gives values between 0.7 and 1.3$M_\odot$.The errors for these mass values are computed as in  \cite{Santos2013} by means of a Monte Carlo analysis. For each case 10 000 random values of effective temperature, surface gravity, and stellar metallicity were drawn from a Gaussian distribution. From the resulting mass distribution, the central value for the mass and 1-sigma uncertainty were derived.
\begin{table*}[h]
\caption{Targets. Spectra source. Stellar Parameters (T$_{\rm eff}$, log~g and [Fe/H]) and errors obtained by FAMA and FASMA for each star of the analysed sample.}
\begin{center}

\begin{tabular}{ll|llllll|llllll}
\hline
\hline
  &  &  &  &  & FAMA &  &  &  &  &  & FASMA &\\
\hline
Star  & Spectra & T$_{\rm eff}$ & eT$_{\rm eff}$ & log~g & elog~g & [Fe/H] &
e[Fe/H]& T$_{\rm eff}$ & eT$_{\rm eff}$ & log~g & elog~g & [Fe/H] & e[Fe/H]\\
            &     & (K) & (K) & (dex) & (dex) & (dex) & (dex) & (K) & (K) & (dex) & (dex) & (dex) & (dex)\\

\hline
55Cnc & UVES & 5358 & 150 & 4.44 & 0.20 & 0.49 & 0.01 & 5295 & 20 & 4.455 & 0.03 & 0.359 & 0.01\\
CoRoT-10 & HARPS & 4908 & 150& 4.31 & 0.20 &0.26 & 0.01& 4972 & 10& 4.439 & 0.03& 0.262& 0.01\\
CoRoT-19 & HARPS & 6331& 150 & 4.07 & 0.20 & 0.15
&0.06 & 6313 & 40& 4.283  & 0.06 & 0.148 & 0.02 \\
\hline
\hline
\end{tabular}

\label{tab:Summary}
\end{center}
\end{table*}
\subsection{FAMA analysis}
\label{sec:2.2}
The aim of FAMA is to allow the computation of the atmospheric parameters and
abundances of a large number of stars using measurements of equivalent widths
(EWs) as automatic and as independent of any subjective approach as possible.
It is based on the simultaneous search for three equilibria: excitation
equilibrium, ionization balance, and the relationship between $\log n$(FeI) and EW/$\lambda$. FAMA also evaluates the statistical errors on individual element
abundances and errors due to the uncertainties in the stellar parameters. The
convergence criteria are not fixed a priori but are based on the quality of the
spectra. The code is described in \cite{Magrini2013}.
For our work, first, we have measured the EWs with DOOp (\emph{DAOSPEC Option Optimiser}, \cite{Cantat-Guadin2014}) an automatic tool developed within the Gaia-ESO survey. The code is based on DAOSPEC code \cite{Stetson2008} and uses Gaussian fit to measure EWs.
We adopted the line list of the Gaia-ESO survey \cite{Heiter2015}, which includes atomic parameters (log~g and damping coefficient) for
a large number of lines in the spectral range 4200-6800 \AA.
We adopted the MARCS model atmospheres (plane parallel and spherical) \cite{Gustafsson2008}.

\subsection{FASMA analysis}
\label{sec:2.3}
The analysis with FASMA is based on the spectral synthesis technique wrapped around the radiative transfer code, MOOG (version 2019; \cite{Sneden1973}).\\
FASMA creates synthetic spectra on-the-fly and delivers the best-fit parameters (effective temperature, surface gravity, iron abundance, and projected rotational velocity) after a non-linear least-squares fit (Levenberg-Marquardt algorithm). The line list is mainly comprised of iron lines initially obtained from VALD3 \cite{Ryabchikova2015}. The regions of the spectral synthesis are defined within small intervals around these iron lines ($\pm$2\AA). The atomic data are calibrated to match the spectra of the Sun and Arcturus but given the strongest constrains on the solar parameters, we gave higher weights to the Sun. The damping parameters are based on the ABO theory \cite{Barklem2000} when available, or in any other case, we use the Blackwell approximation. The model atmospheres are interpolated from the ATLAS grid \cite{Kurucz1993} in LTE. The minimization is a two step procedure where initially we obtain the best-fit values using fixed solar macro- and microturbulence. Then, we refine our results in a second step, with updated macro- and micro-turbulence based on empirical relations. Macroturbulence velocity is set by the calibration of \cite{Doyle2014} and microturbulence is set based on calibrations for either dwarfs \cite{Tsantaki2013} or giants \cite{Adibekyan2015}. The uncertainties are derived from the covariance matrix constructed by the nonlinear least-squares fit. The methodology is described in detail in \cite{Tsantaki2018,Tsantaki2020} where it is tested for both giant and dwarf samples, including stars with larger rotation.
\section{Comparison among the methods}
\label{sec:3}
\begin{figure*}
\includegraphics[width=0.5\textwidth]{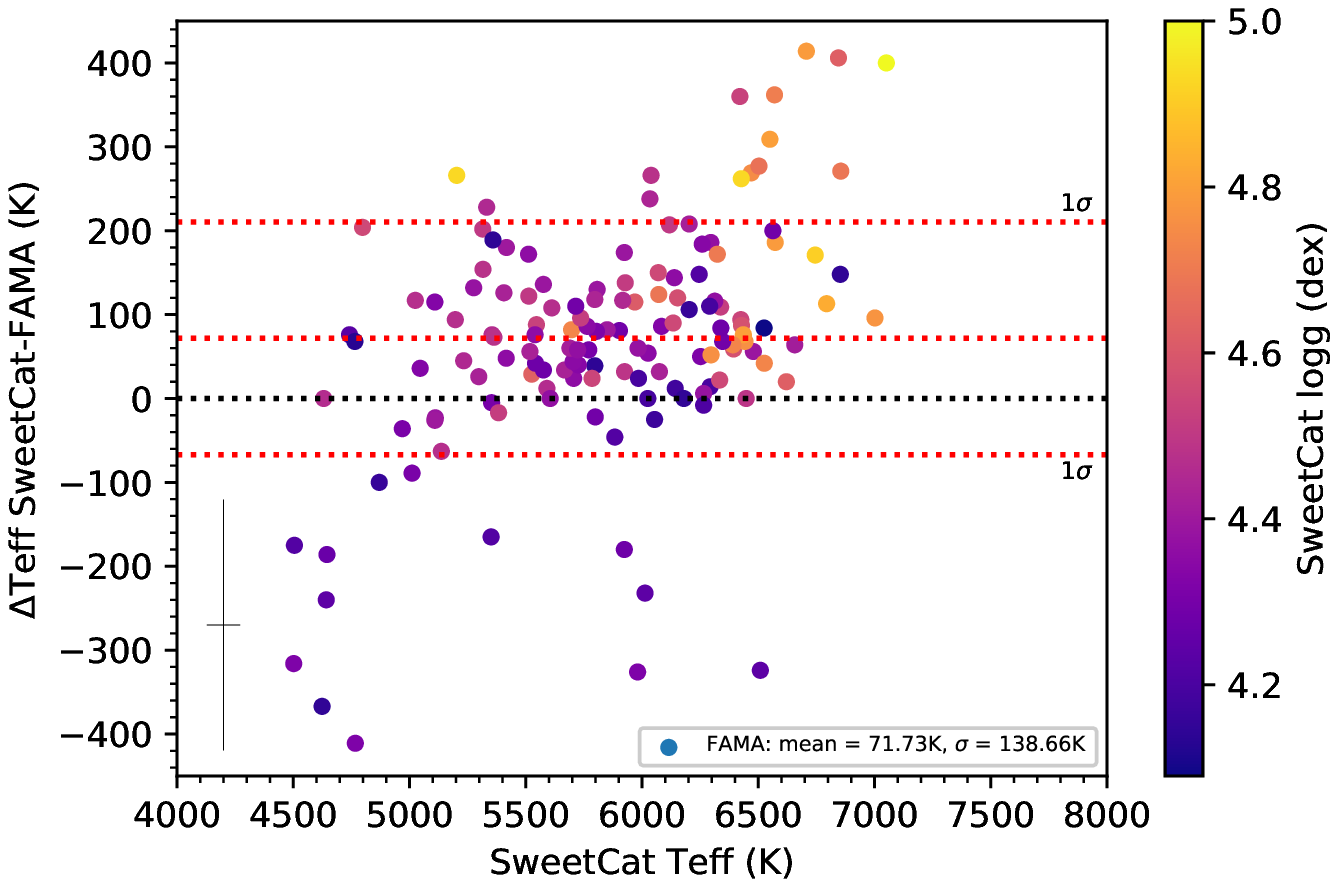}
 \includegraphics[width=0.5\textwidth]{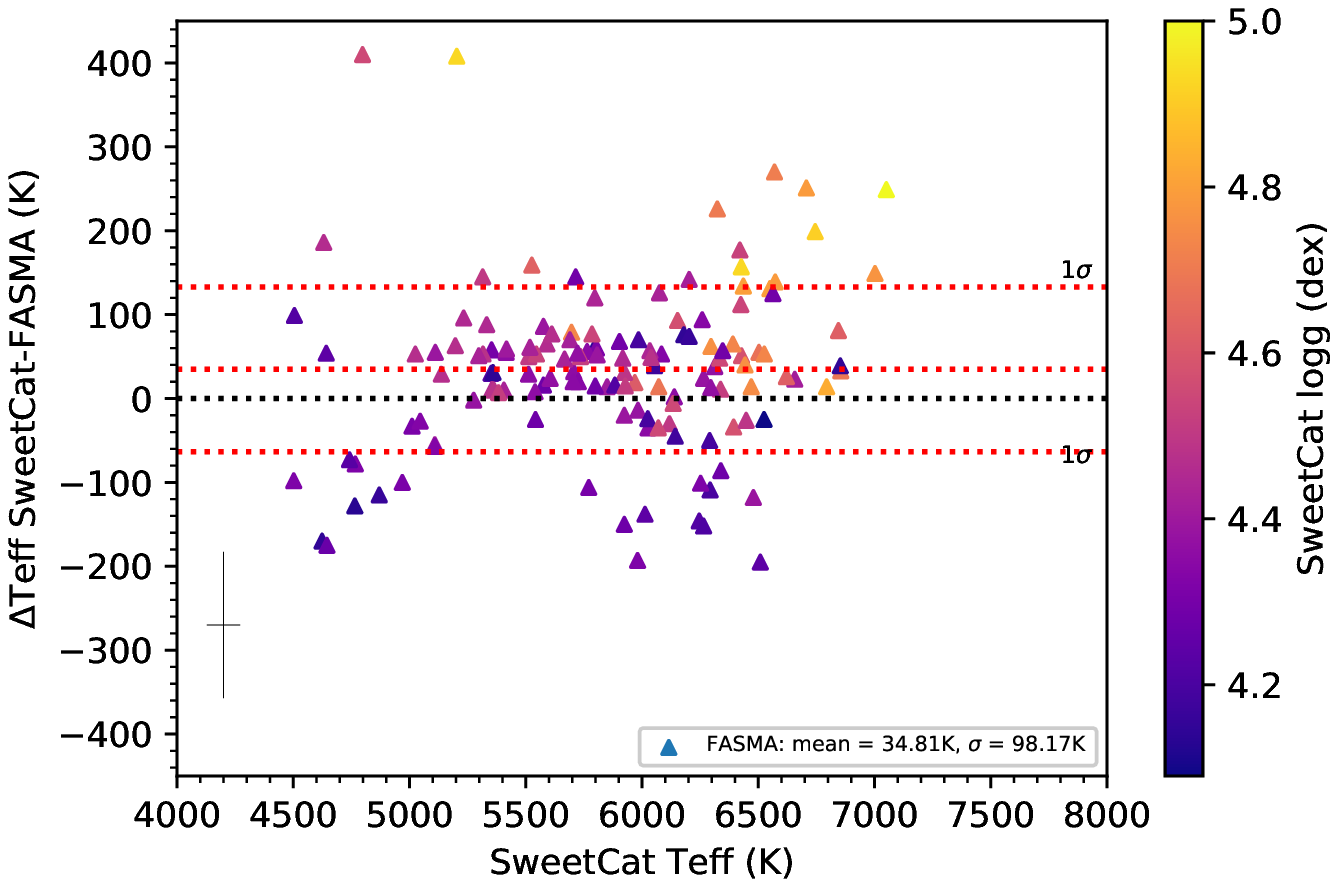}
 \includegraphics[width=0.5\textwidth]{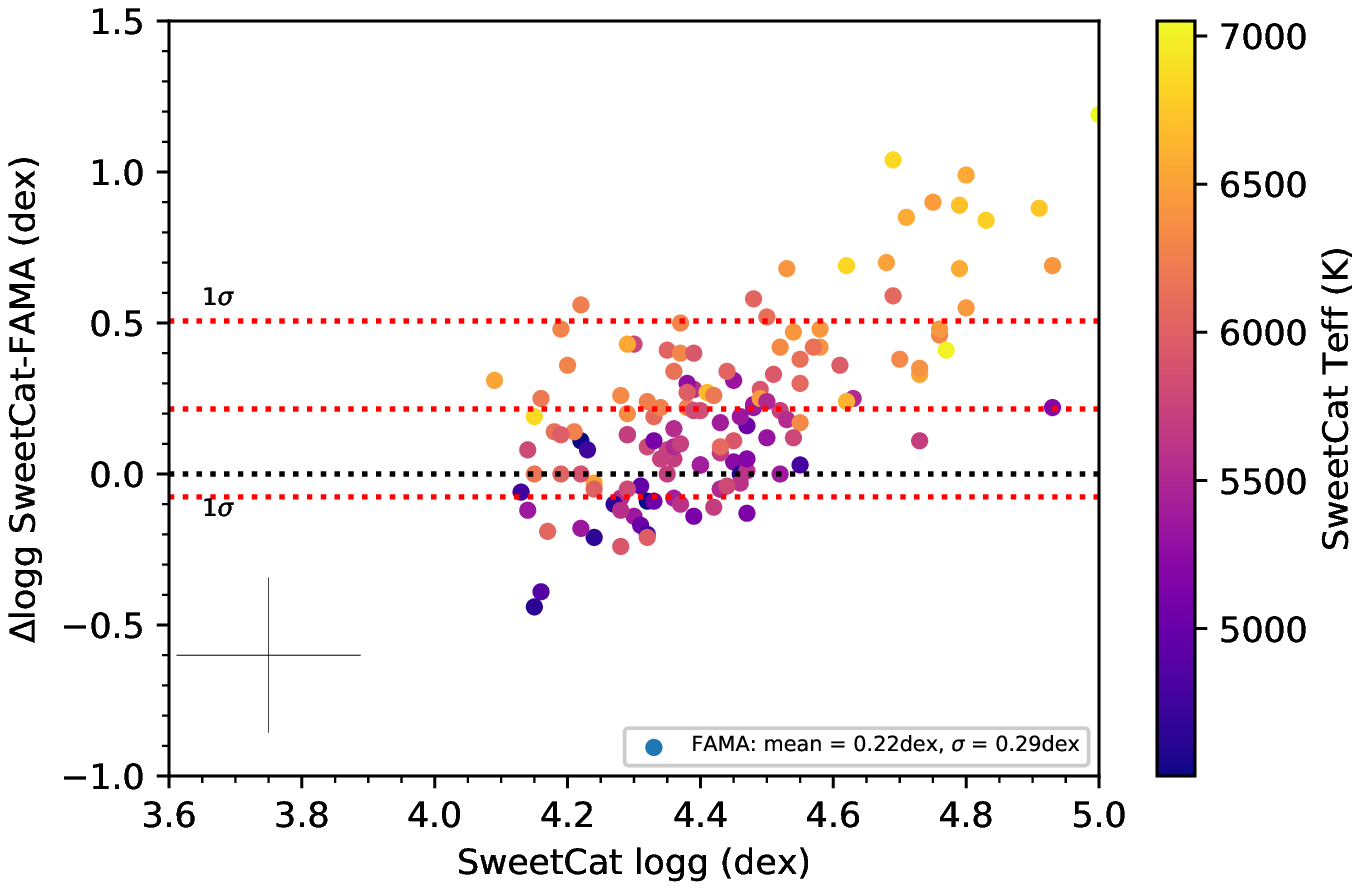}
 \includegraphics[width=0.5\textwidth]{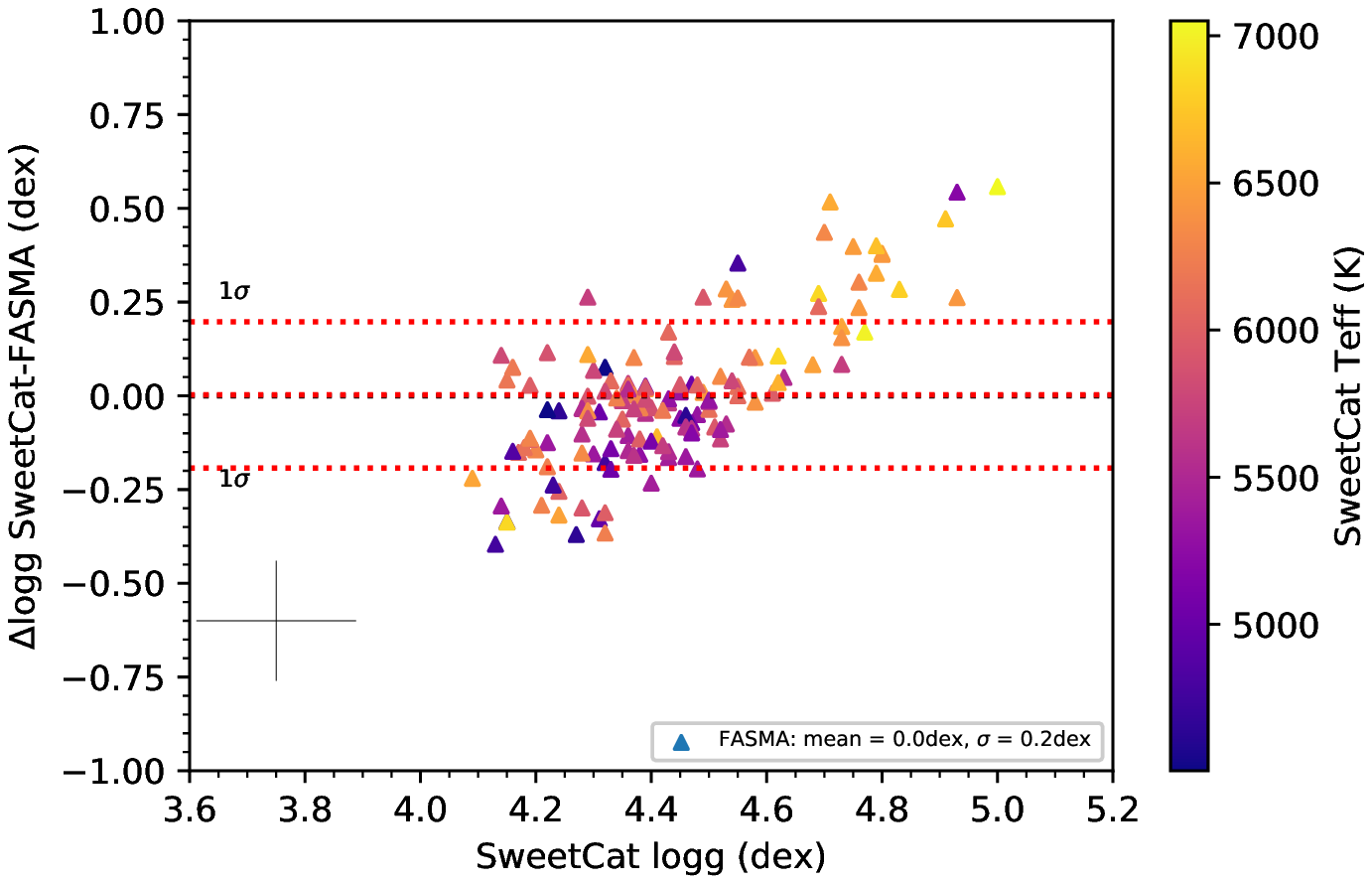}
 \includegraphics[width=0.5\textwidth]{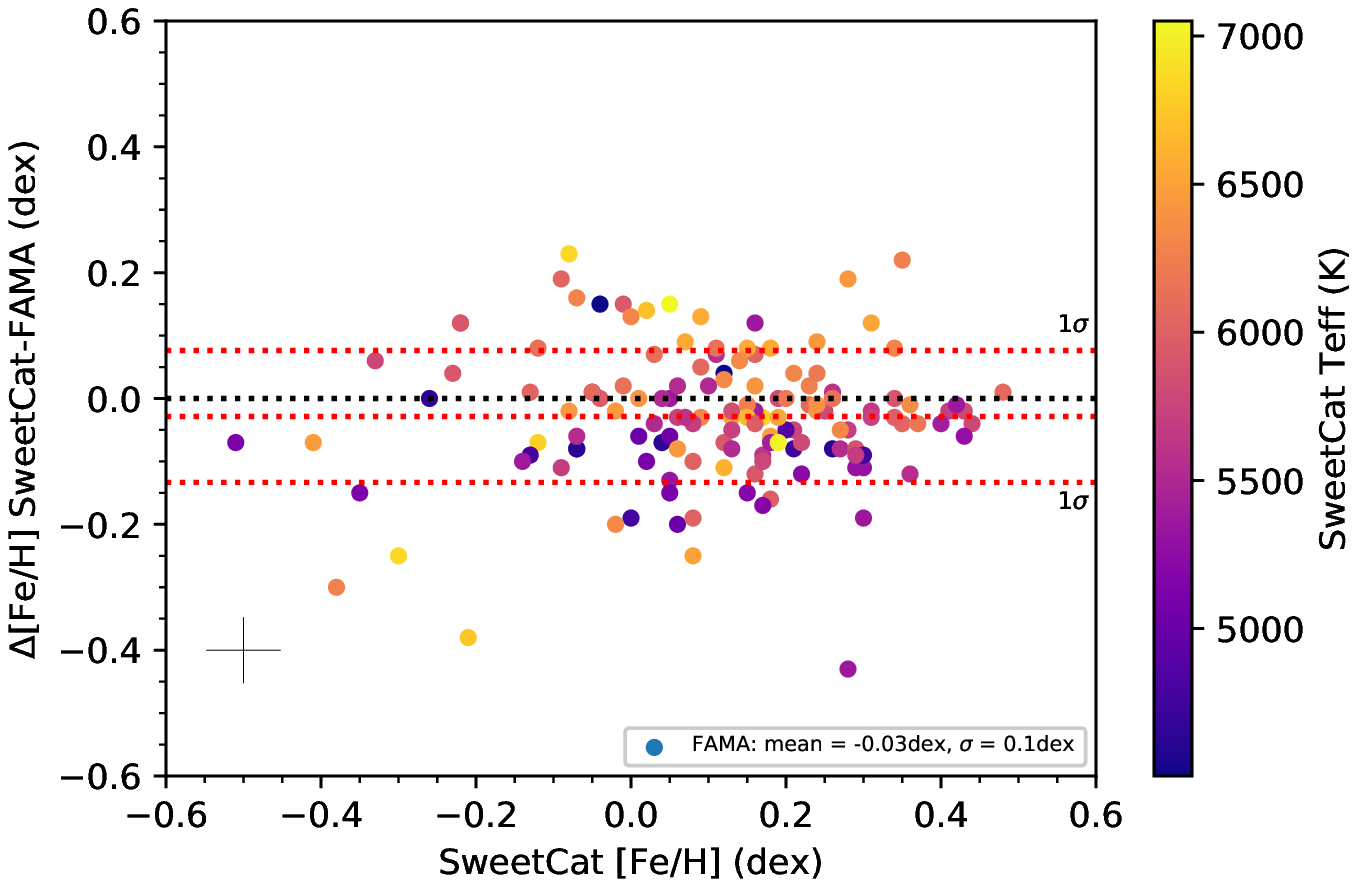}
 \includegraphics[width=0.5\textwidth]{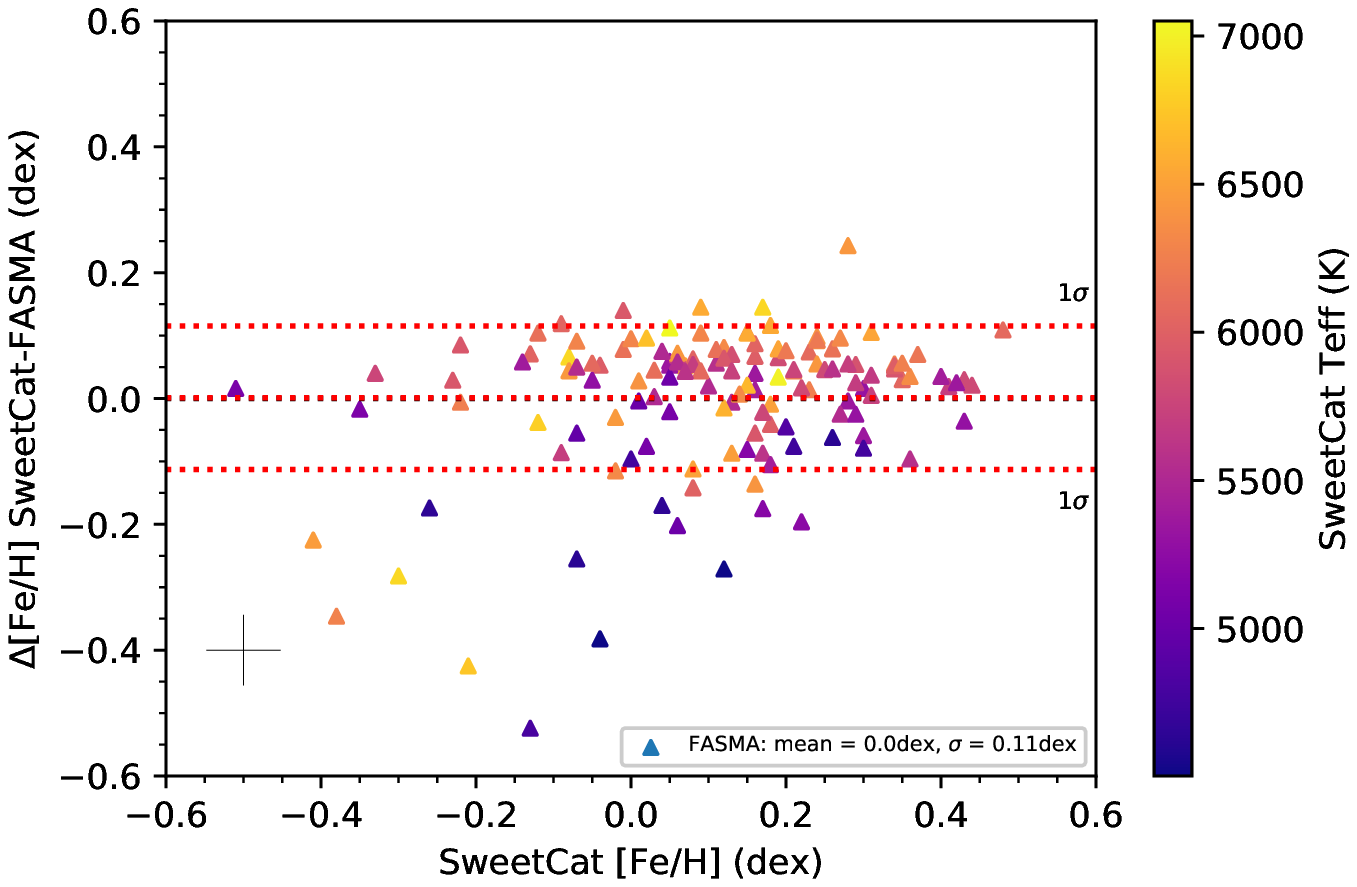}
\caption{Top panels: Difference between T$_{\rm eff}$ from SWEET-Cat and T$_{\rm eff}$ from FAMA (left)/FASMA (right) as a function of the T$_{\rm eff}$ SWEET-Cat. The colour scale indicates the value of log~g for SWEET-Cat. Red dashed line: average of the T$_{\rm eff}$ difference and 1$\sigma$ level.
Central panels: Difference between log~g from Sweet-Cat and log~g from FAMA (left)/FASMA (right) as a function of log~g from SWEET-Cat.The colour scale indicates the value of T$_{\rm eff}$ for SWEET-Cat.
Bottom panels: Difference between [Fe/H] derived using SWEET-Cat and FAMA (left)/FASMA (right)  as a function of [Fe/H] from SWEET-Cat. The colour scale represents T$_{\rm eff}$ from SWEET-Cat. Average errors are also indicated on the left of each plot.}
\label{fig:2}       
\end{figure*}
We compare our results seeking possible outliers and trends among the parameters obtained by the three methods.
In this preliminary work, we explore the results of the three analysis, comparing with external benchmarks, as the surface gravities derived from light-curves in literature and from {\it Gaia} parallaxes and photometry, to identify the ranges of best performances. 

\subsection{SWEET-Cat vs FAMA/FASMA: T$_{\rm eff}$}
Fig.~\ref{fig:2} (top panels) shows the difference in effective temperature (T$_{\rm eff}$) between SWEET-Cat and FAMA (left panel) or SWEET-Cat and FASMA (right panel) as a function of the SWEET-Cat T$_{\rm eff}$. The color scale represents log g for SWEET-Cat.

The bulk of the T$_{\rm eff}$ results for the SWEET-Cat and FAMA is in good agreement (within 1-$\sigma$), although there are some outliers: 28 stars show a difference in T$_{\rm eff}$ over 1$\sigma$ level but still within $\sim$417K (3$\sigma$ level). Among these outliers, we noticed that 7 stars present large vsini ($>$10km/s).
The T$_{\rm eff}$ spread seems more evident when considering stars with the lower and higher T$_{\rm eff}$ in the considered range. In addition, there is an offset between the
two methods, with the mean value (red dashed central line) for the difference of $\sim$70 K.\\
Focusing now on the FASMA analysis, also in this case, the bulk of the T$_{\rm eff}$ results for the SWEET-Cat and FASMA is in good agreement (within 1$\sigma$), although outliers are also present. For stars with the lower and higher T$_{\rm eff}$ in the range of interest, the dispersion between the two methods increases. 
An offset between the two methods is present with a mean value (red dashed line) in the difference of $\sim$34K.
We note here that some of the outliers are not in common with the outliers found using the FAMA method.
\subsection{SWEET-Cat vs FAMA/FASMA: log g}
Fig.~\ref{fig:2} (central panels)  show respectively the comparison of log g derived using the FAMA (left panel) or FASMA (right panel) methods vs the results given in SWEET-Cat.
The colour scale represents T$_{\rm eff}$ for SWEET-Cat. 
There are trends in both the FAMA and FASMA results: the differences between their log~g
and those of SWEET-cat are increasing with log~g. The differences are particularly
high for higher surface gravity (log~g $> $4.6\,dex). In the log~g range of $\sim4-4.6$\,dex, the trend is almost negligible. 
In addition, FAMA, on average, gives lower gravities, of about $\sim-0.18$\,dex.\\
\subsection{SWEET-Cat vs FAMA/FASMA: [Fe/H]}
Fig.~\ref{fig:2} (bottom panels)  show respectively the comparison of [Fe/H] derived using the FAMA (left panels) method and FASMA (right panels) vs the results given in SWEET-Cat. The colour scale represents T$_{\rm eff}$ for SWEET-Cat.

Considering the typical errors on metallicity ($\sim$0.10 dex), the results are quite
satisfactory: there are almost no trend in the differences vs [Fe/H] and the
means are close to zero (slightly negative for FAMA).

This is one of the most important results of the analysis and we need to remark it.
Despite the differences in the derived stellar parameters, the three methods converge to very similar final metallicities [Fe/H]. 
\begin{figure*}[h]
\includegraphics[width=0.5\textwidth]{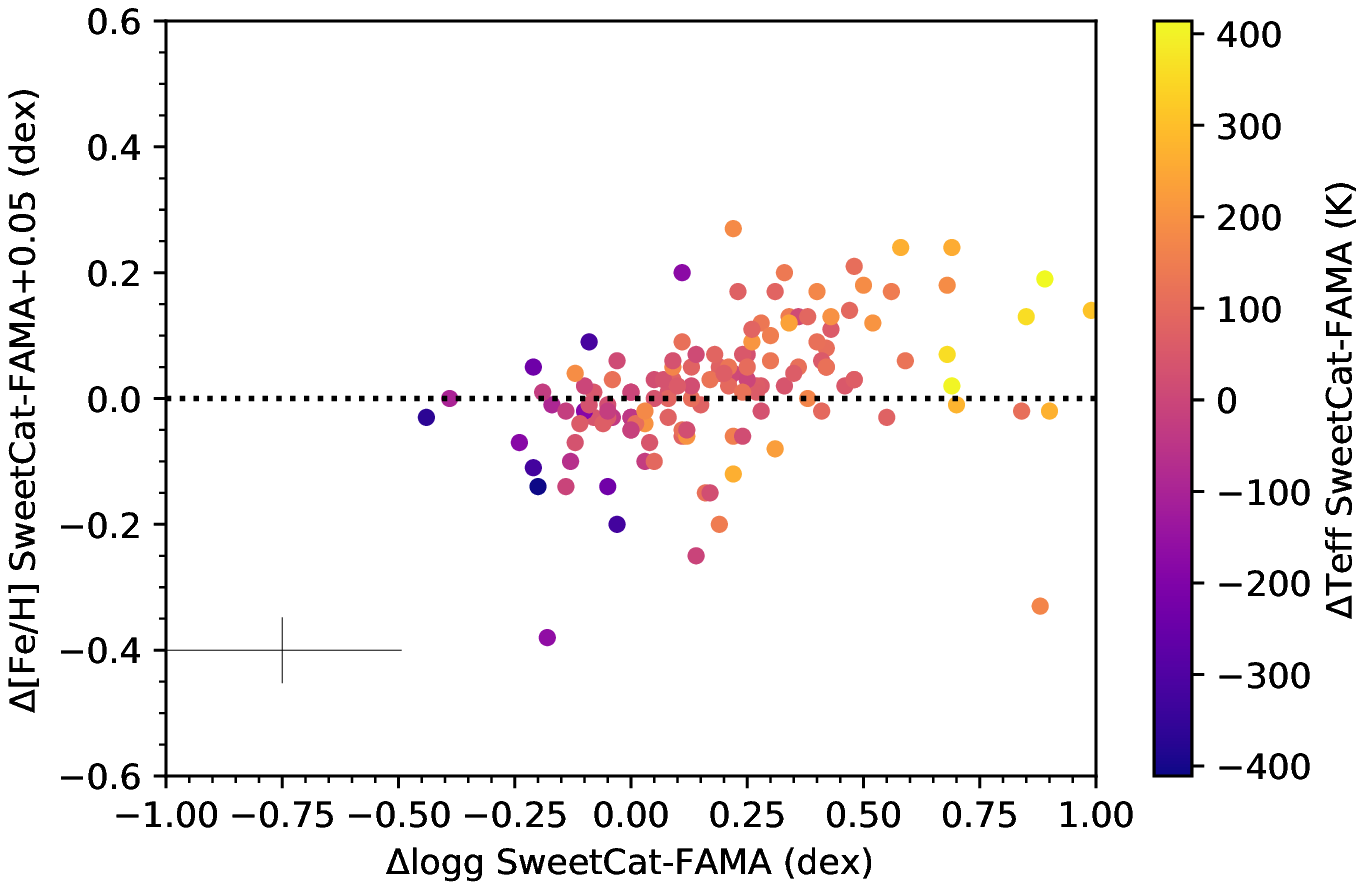}
\includegraphics[width=0.5\textwidth]{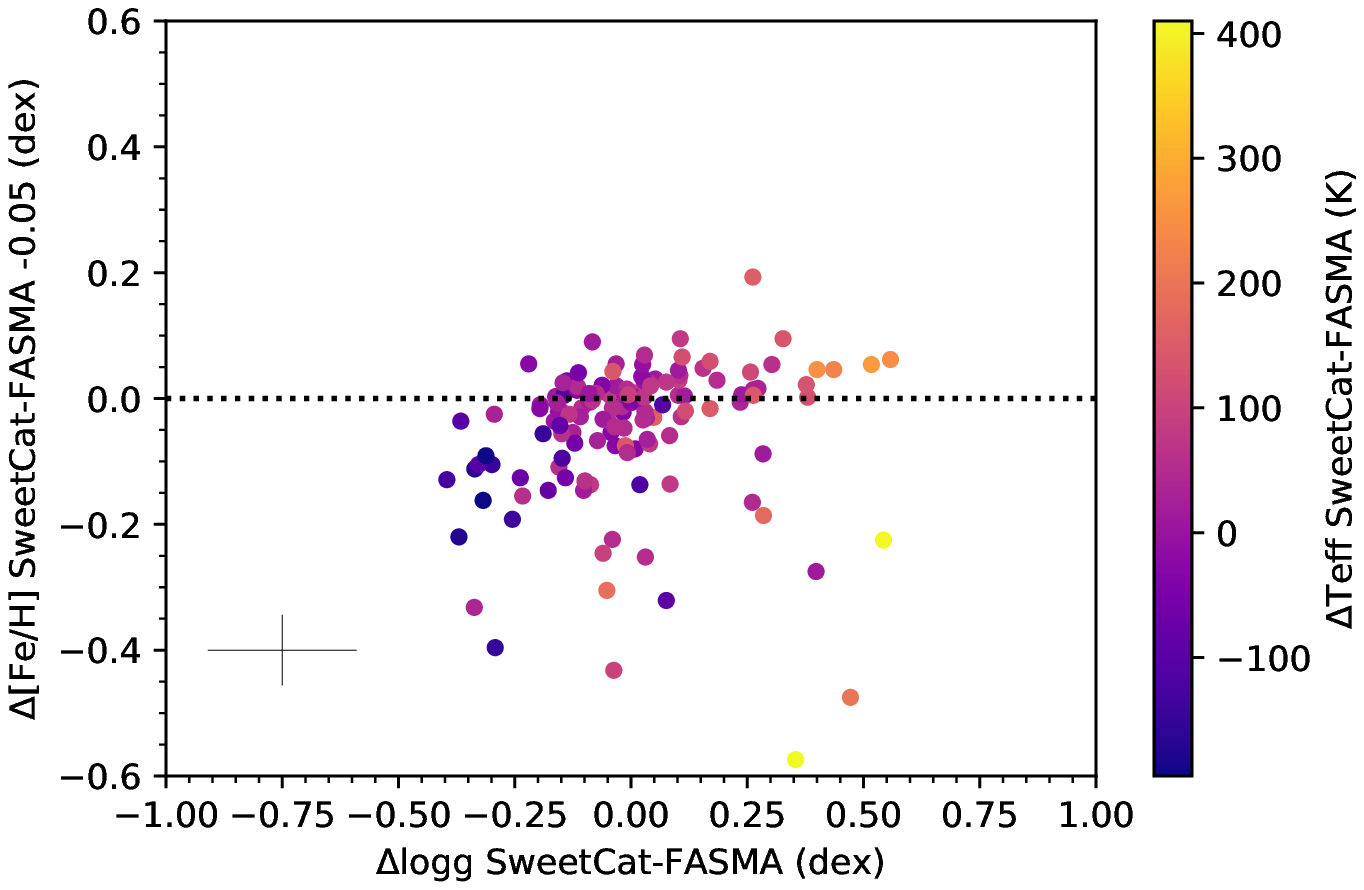}
\caption{Left panel: SWEET-Cat vs FAMA differences in log~g vs differences in [Fe/H], colour-coded by differences in T$_{\rm eff}$. Right panel: the same plot for the FASMA results. An offset of +0.05~dex is applied to FAMA metallicities, while an offset of -0.05~dex to the FASMA ones.   }
\label{fig:3}  
\end{figure*}
The three methods rely, indeed, on different line lists, with different set of atomic data, which make them to converge on slightly different sets of T$_{\rm eff}$ and log~g values, which should satisfy the conditions of excitation equilibrium and ionization balance.  T$_{\rm eff}$ and log~g tend to vary along a local minimum, in which the value of [Fe/H] is usually more constant.

In Fig.~\ref{fig:3}, we show the differences in metallicity of the FAMA and FASMA results with respect to SWEET-Cat, as a function of the differences in the derived surface gravities. The symbol are colour-codes by the differences in the derived T$_{\rm eff}$, again, with respect to the SWEET-Cat ones. 

Removing the systematic offsets (+0.05~dex from FAMA and -0.05~dex for FASMA with respect to the SWEET-Cat scale), the differences in [Fe/H] are negligible for variation in the surface gravity within $\pm$0.3~dex with respect to the SWEET-Cat ones. In the comparison between FAMA and SWEET-cat [Fe/H] even for differences in log~g$\sim$ -0.3~dex, the agreement in metallicity between the two methods is still good. 
Larger differences in log~g$>$+0.3~dex, corresponds to higher [Fe/H] in SWEET-Cat than the ones obtained with the other methods.
This means that moving to higher effective temperatures the discrepancies between the methods increases also for the results in [Fe/H] and
the hottest temperature regime remains thus critical also to obtain reliable metallicities.\\
Concluding, the stars in the parameter space closer to the solar values
(T$_{\rm eff}$=$5000-6000$K, log~g=$4.2-4.6$ dex) are those for which the three methods are in better agreement.
Among the three parameters, the best accord is reached for [Fe/H] in the whole parameter space. The most critical parameter is, however, the surface gravity: external comparison (using {\it Gaia} data, asteroseismic data, isochrones) are needed to evaluate the results of the different methods.
A correct evaluation of all stellar parameters is indeed fundamental for a precise characterisation of the host star, including the determination of its age \cite{Bossini2020}.

\begin{figure*}
 \includegraphics[width=0.5\textwidth]{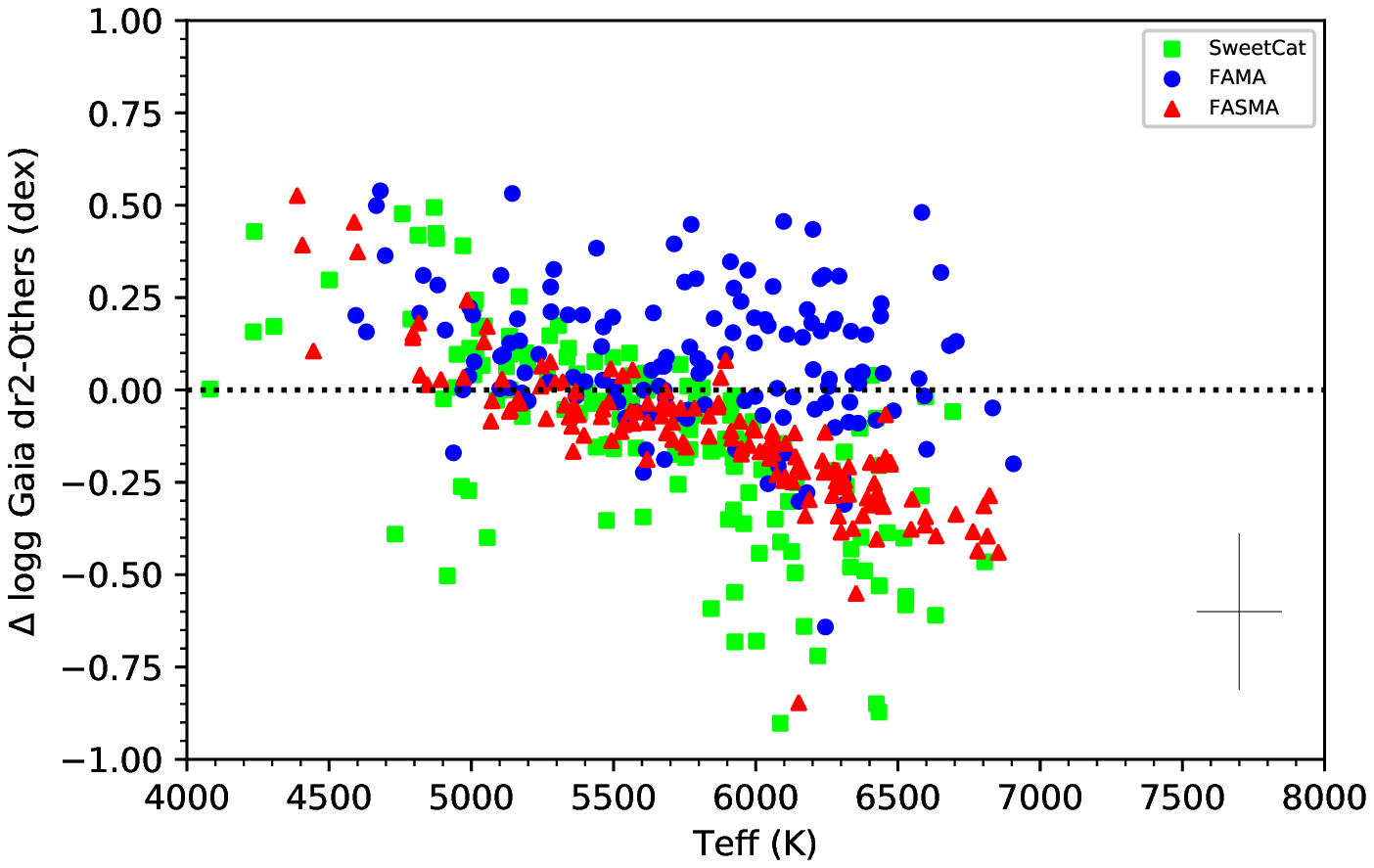}
 \includegraphics[width=0.5\textwidth]{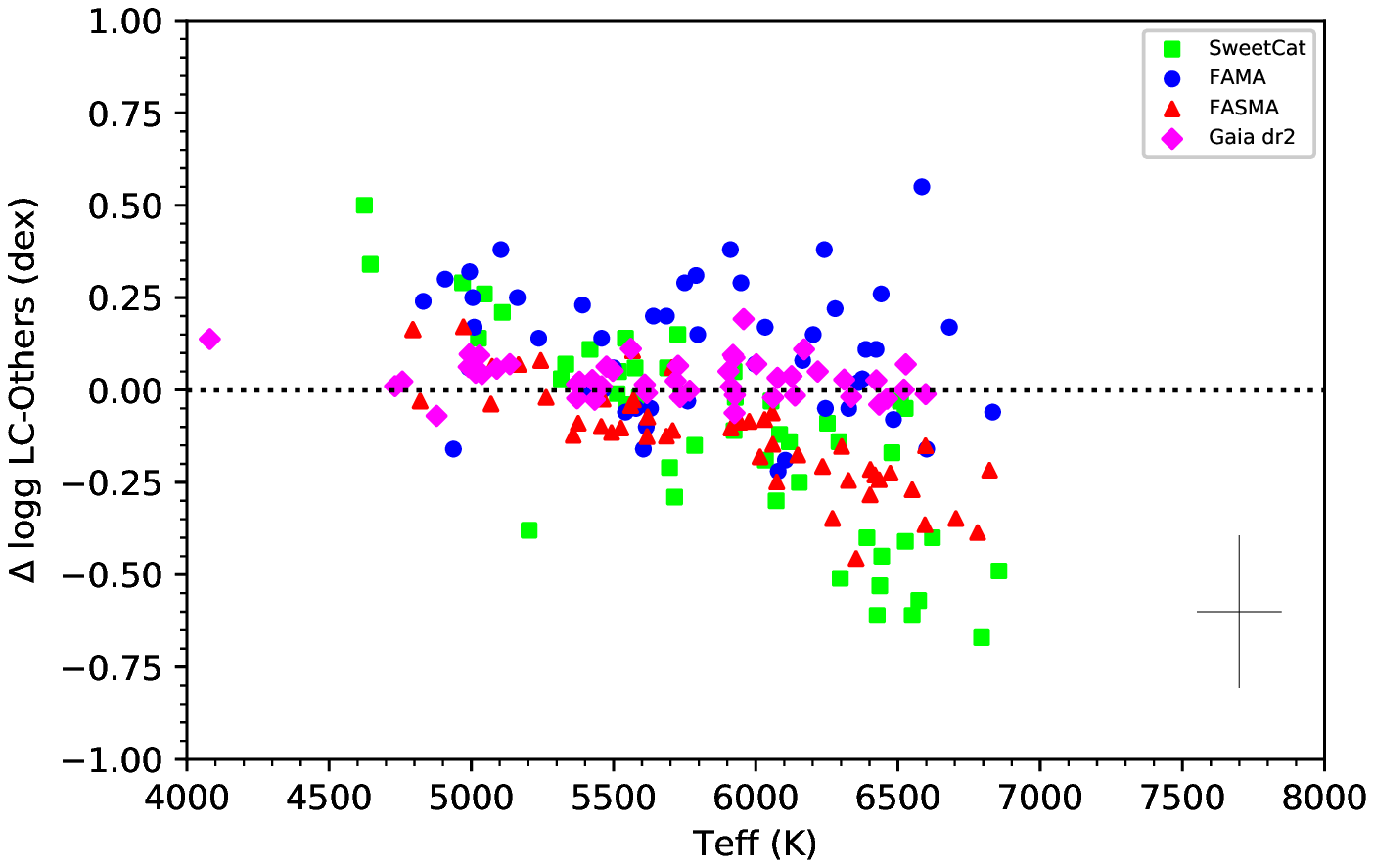}
 \includegraphics[width=0.5\textwidth]{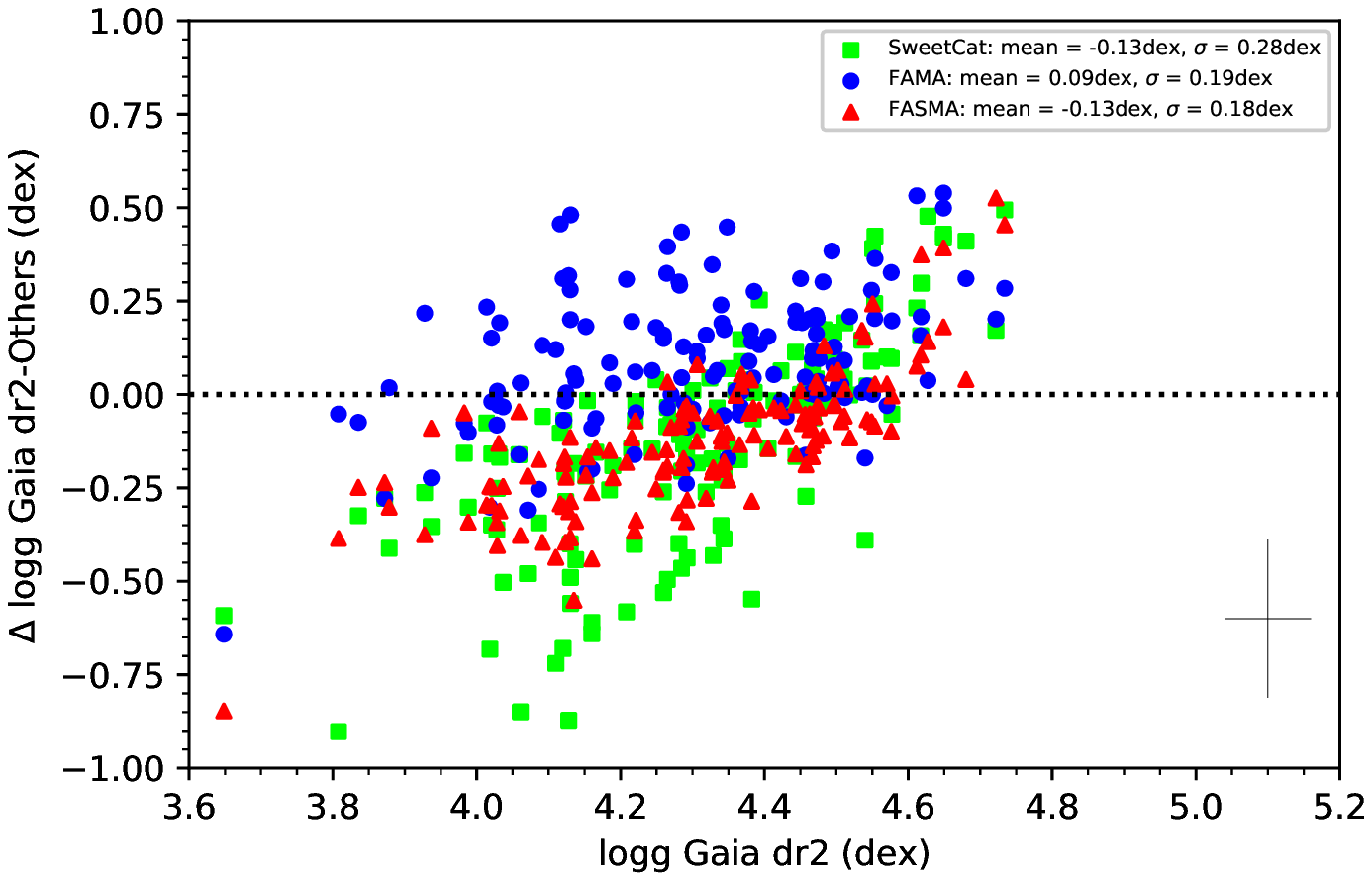}
 \includegraphics[width=0.5\textwidth]{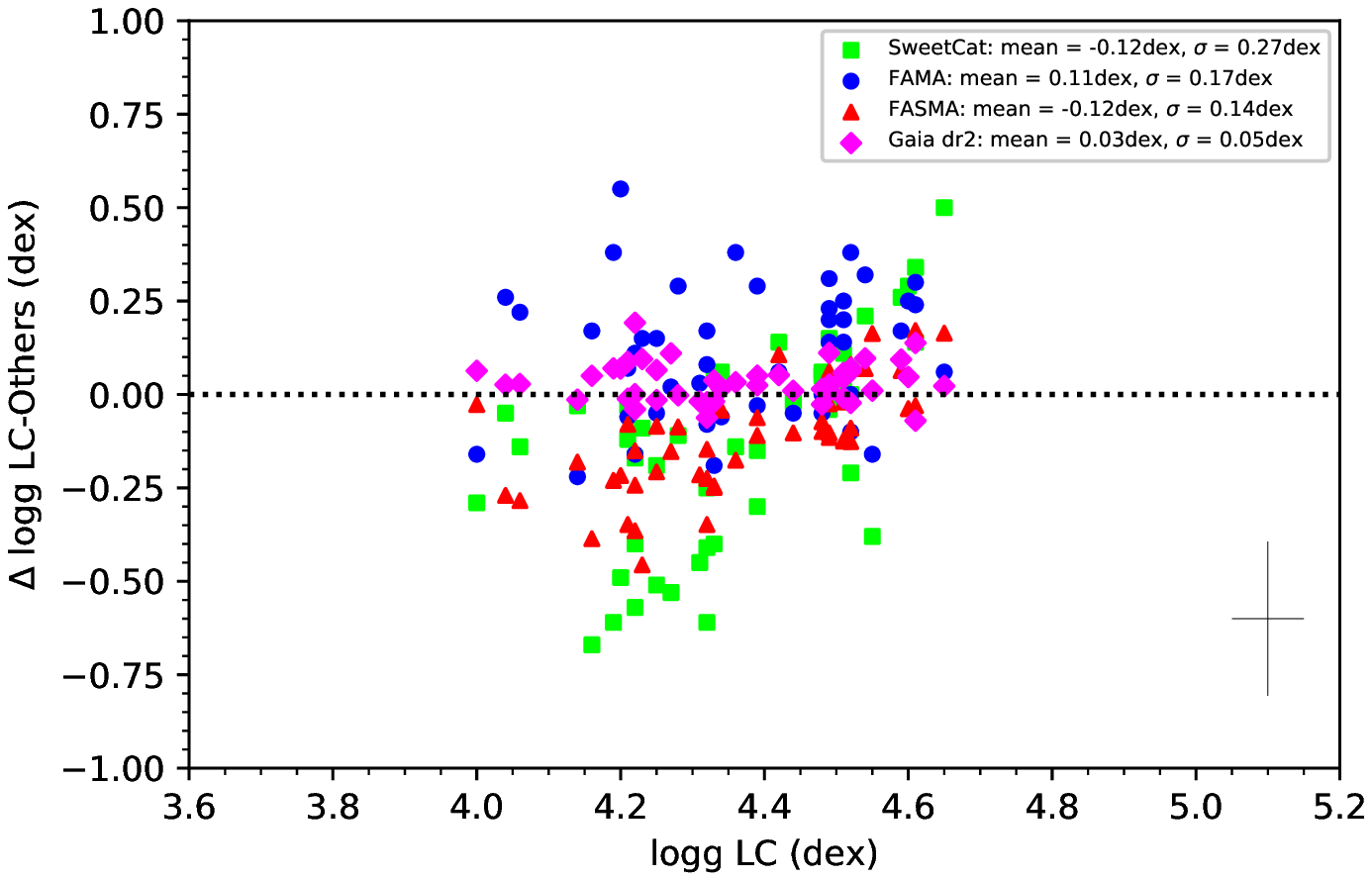}
\caption{Left panels: Comparison between spectroscopic (Sweet-cat, FAMA and FASMA) and trigonometric log~g as a function of T$_{\rm eff}$ (top) and of trigonometric log~g (bottom). The corresponding T$_{\rm eff}$ for the trigonometric log~g, is derived from the {\it Gaia} measurements. Right panels: Comparison between spectroscopic (Sweet-cat, FAMA and FASMA) and log~g derived from the light-curves as a function of T$_{\rm eff}$ (top) and of light-curve log~g (bottom). Average errors are also indicate on a side of each plot. For the parameter differences we considered as error the larger value among the average errors from each method.}
\label{fig:4}       
\end{figure*}

\section{Evaluation of the accuracy and precision of the stellar parameters}
\label{sec:4}
In this section, we discuss some indirect checks on the accuracy and precision of the stellar parameters.
We especially focus on the control of the stellar surface gravity, parameter that cannot be usually constrained well by spectroscopic methods (e.g. \cite{Mortier2013,Tsantaki2014}).  This fact has an impact on the calculation of other stellar parameters (T$_{\rm eff}$ and [Fe/H]) and subsequently on the derivation of the stellar chemical abundances. 

\subsection{Comparison with surface gravity derived from trigonometric distances using {\it Gaia} {\sc dr2}}
\label{sec:4.1}
We compare the log~g derived using the three spectroscopic methods with the
trigonometric log~g based on {\it Gaia} {\sc dr2} photometry and parallax. 
Photometric gravities have been obtained using the following equation
\begin{equation}
\log(g) = log(M/M_{\odot})+0.4\cdot M_{bol}+4\cdot \log(T_{\rm eff})-12.505 
\end{equation}
where $M/M_{\odot}$ is the stellar mass (in solar mass units) provided by the Sweet-cat database. $M_{\rm bol}$ is the bolometric magnitude, obtained from the luminosity published in the {\it Gaia} {\sc dr2} catalog \cite{Gaia2018}
using the following relation M$_{\rm bol}$=4.75-2.5$\times\log$(L/L$_{\odot}$), and  T$_{\rm eff}$ is the {\it Gaia} photometric T$_{\rm eff}$.  
Since the stars in our sample are nearby and thus the reddening is negligible, we can derive $M_{bol}$ from the {\it Gaia} luminosities and the distances obtained directly by inverting the parallaxes.
In Fig.~\ref{fig:4} (left top and bottom panels) we show the comparison (no outliers are removed in the plot). 
Surface gravities from the SWEET-Cat appear to be slightly underestimated at low T$_{\rm eff}$, and highly overestimated at high T$_{\rm eff}$. The
same behaviour for FASMA at high T$_{\rm eff}$. No clear trend appears for the FAMA surfaces gravities vs T$_{\rm eff}$.
\begin{table*}[h]
\caption{Mean differences log~g$_{\it Gaia}$ - log~g$_{\rm spec}$  with standard deviation (1-$\sigma$) and median differences (in parenthesis) in three T$_{\rm eff}$ intervals. }
\begin{center}

\begin{tabular}{lllll}
\hline
\hline
Method  &   T$_{\rm eff} <$5000~K       & 5000~K$<$T$_{\rm eff} <$6000~K     & T$_{\rm eff} >$ 6000~K    &   Total T$_{\rm eff}$ range\\
\hline
SWEET-Cat & 0.132$\pm$0.068 (0.165)  &-0.042$\pm$0.017 (-0.032) & -0.381$\pm$0.037 (-0.393) & -0.115$\pm$0.023 (-0.083)\\
FAMA & 0.162$\pm$0.043 (0.160) &  0.059$\pm$0.020 (0.053) & 0.108$\pm$0.029 (0.126) & 0.088$\pm$0.016 (0.064) \\
FASMA &0.088$\pm$0.051 (0.037) &-0.086$\pm$0.011 (-0.084) & -0.280$\pm$0.014 (-0.289)  & -0.137$\pm$0.013 (-0.116) \\
\hline
\hline
\end{tabular}

\label{tab:diff:logg}
\end{center}
\end{table*}
In Table~\ref{tab:diff:logg} we divide the sample in three temperature regimes: T$_{\rm eff}<$5000 K, 5000 K$<$T$_{\rm eff} <$6000 K, and finally T$_{\rm eff}>$ 6000 K. 
We compute the mean differences between the log~g from {\it Gaia} and the spectroscopic log~g from the three methods, and their standard deviation (1-$\sigma$). 
In the coolest regime, all methods tend to underestimate the surface gravities, with differences consistent with each other within the error.
The spectral synthesis method FASMA provides gravities, on average, in better agreement with  the trigonometric ones. 
In the intermediate regime, with 5000 K$<$T$_{\rm eff}<$6000 K, the three methods show the better agreement with the  log~g from {\it Gaia}: FAMA slightly underestimates the trigonometric gravities, while SWEET-Cat and FASMA slightly  overestimate them. 
The most challenging regime is the hottest one, with T$_{\rm eff} >$ 6000 K. 
In this temperature range, there is no agreement between the spectroscopic methods and the trigonometric one (FASMA and SWEET-Cat overestimate the trigonometric log~g, while FAMA slightly underestimates it). 
In the last column, we report the mean differences for the whole sample: slightly negative differences for SWEET-cat and FASMA, and positive for FAMA. However, it is clear that global mean differences mask the most critical regimes, for the hottest and coolest stars of the sample.  

\subsection{Comparison with log~g derived from the light-curves}
\label{sec:4.2}
\begin{figure}
\includegraphics[width=0.50\textwidth]{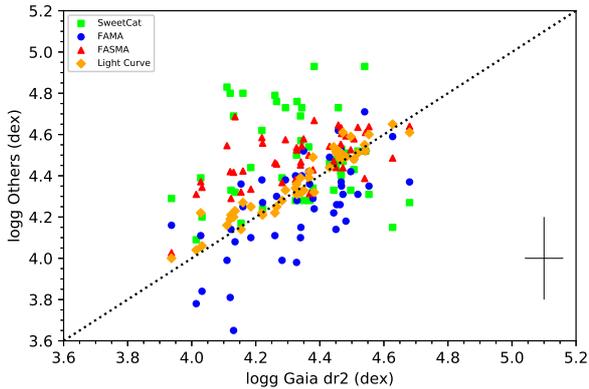}
\caption{Trigonometric surface gravity vs surface gravity derived with the other methods, including spectroscopic methods and light-curve method.}
\label{fig:5}       
\end{figure}
We compare the trigonometric and spectroscopic gravities with those derived from the light-curves in literature, and available in the SWEET-Cat database.
In Fig.~\ref{fig:4} (right top and bottom panels)  we show the difference between the log~g derived from the light-curves (literature values for 48 targets were found) and the respective spectroscopic log~g listed in the SWEET-Cat (green filled squares) and obtained through FASMA (red filled triangles) and FAMA (blue filled circles) as a function of T$_{\rm eff}$ and of the light-curves log~g. 
First, we notice the good agreement between the trigonometric and light-curve surface gravities,
despite the two methodologies are quite different, and this is an encouraging result (see also Fig.~\ref{fig:5}). 
Concerning the other methods, similar considerations as in the previous section can be obtained: 
FAMA shows a bit higher dispersion and a positive offset, but almost no trend. FASMA displays the lower dispersion, but tends to
overestimate log~g at high T$_{\rm eff}$. SWEET-cat seems to overestimate log~g at high T$_{\rm eff}$, and underestimate at low T$_{\rm eff}$.
Clearly the hottest region is the most critical one, both because hot stars have less absorption lines useful to constrain the photospheric parameters and because stellar rotation might become important, making more difficult to measure absorption lines which can be blended. 
Moreover, in the low-temperature regime, the presence of molecular bands can blend and hide the atomic lines.

\subsection{Comparison with isochrones }
\label{sec:4.3}

\begin{figure*}
 \includegraphics[width=0.5\textwidth]{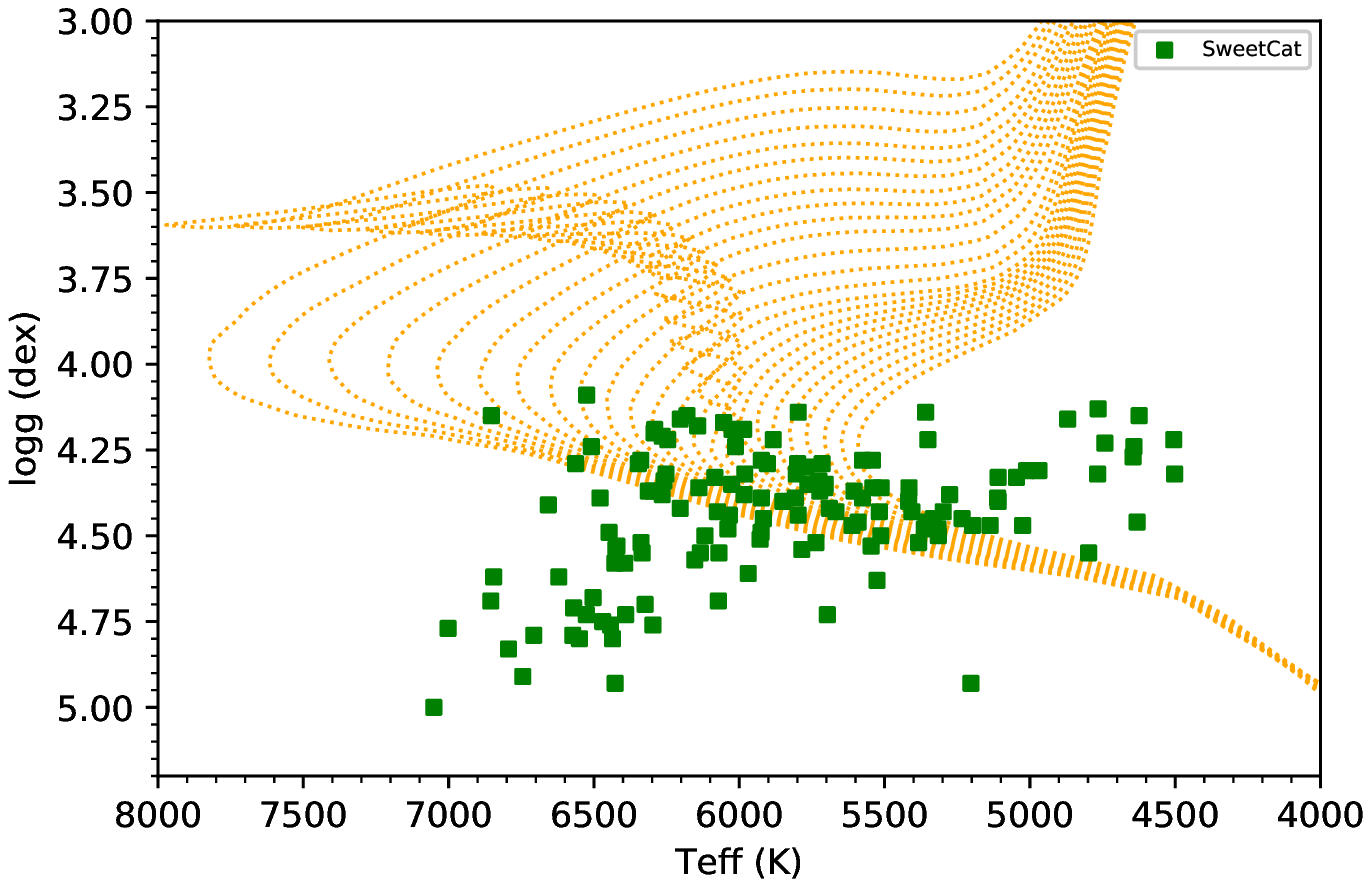}
 \includegraphics[width=0.5\textwidth]{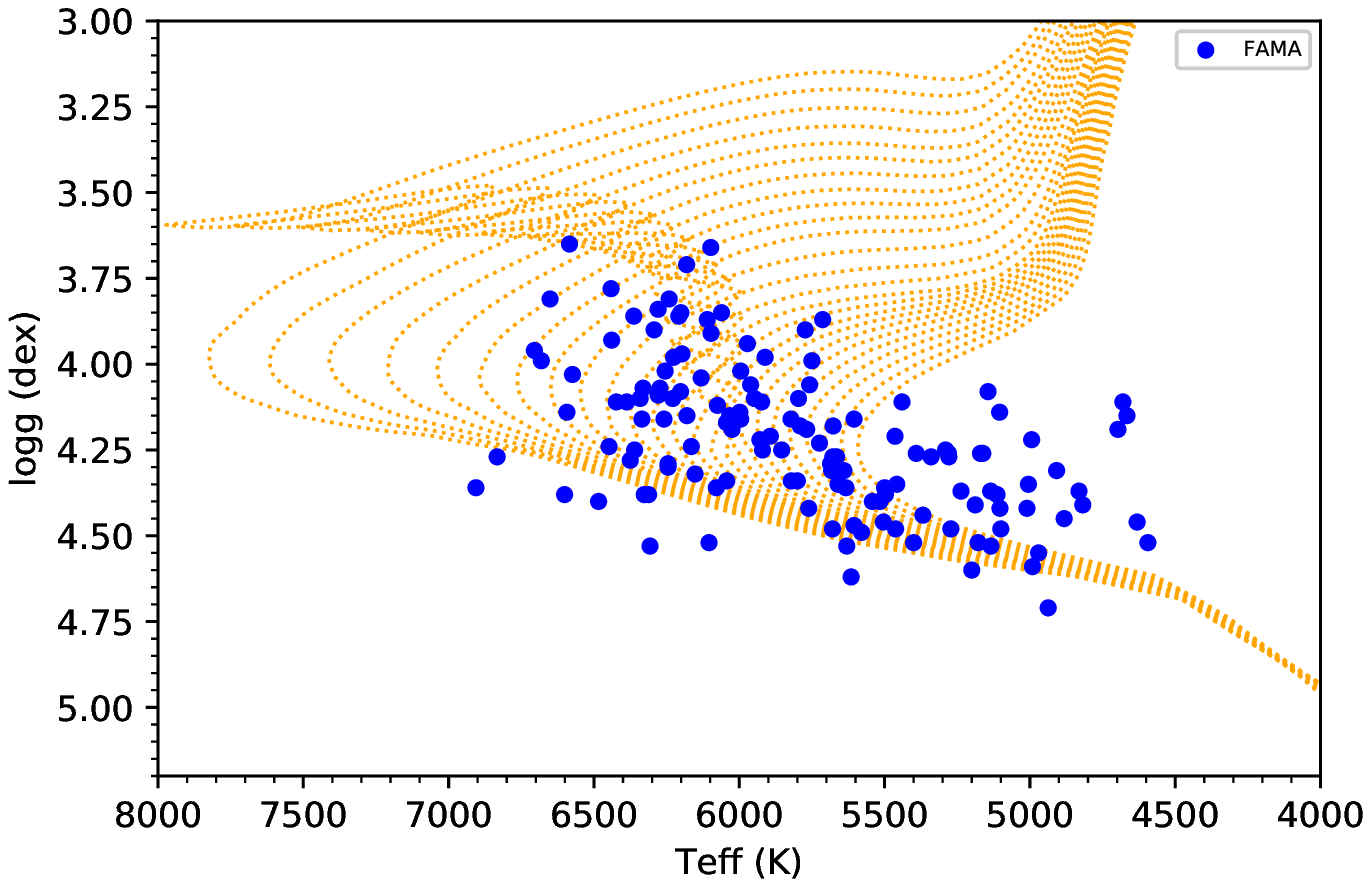}
 \includegraphics[width=0.5\textwidth]{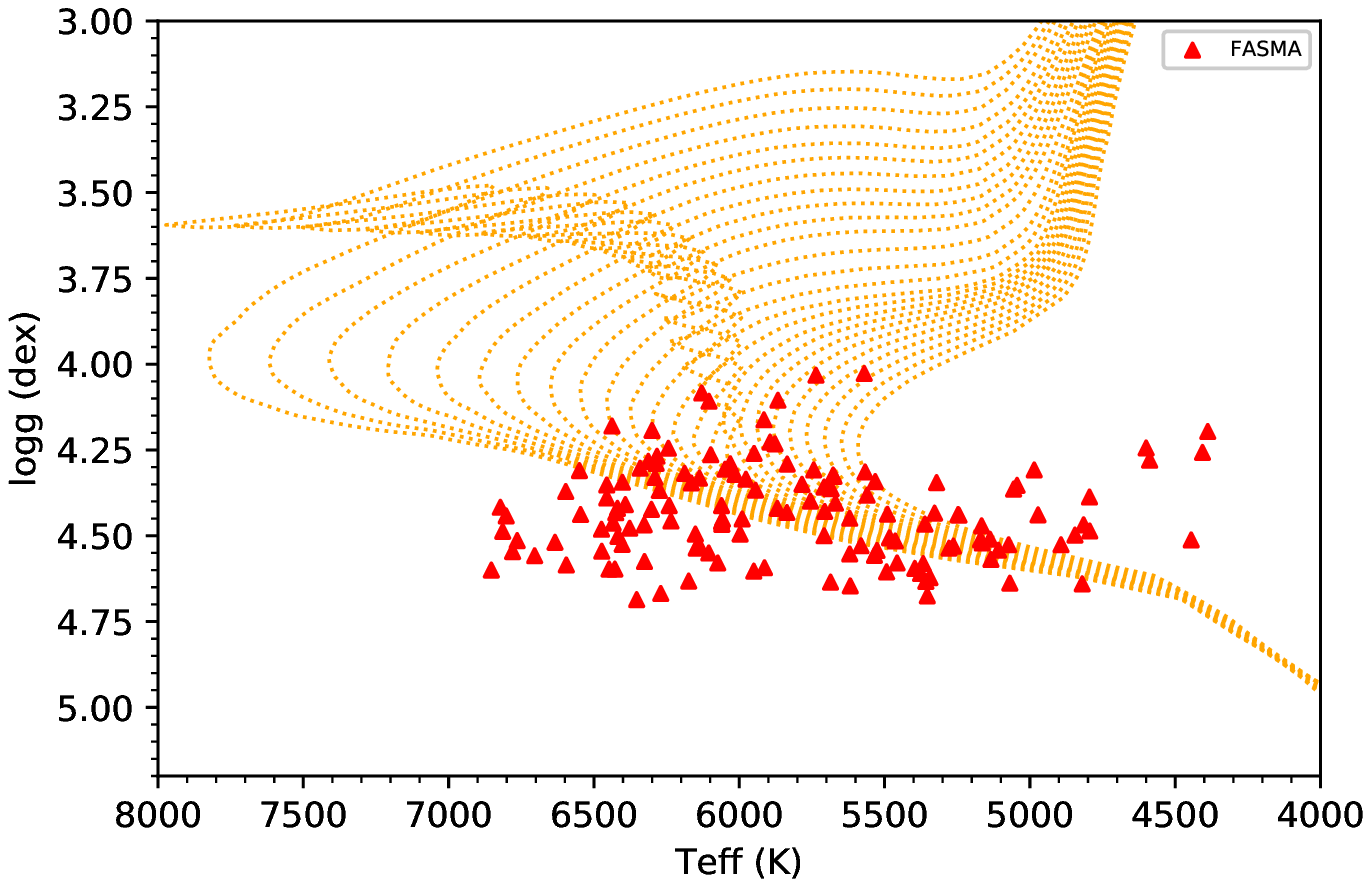}
 \includegraphics[width=0.5\textwidth]{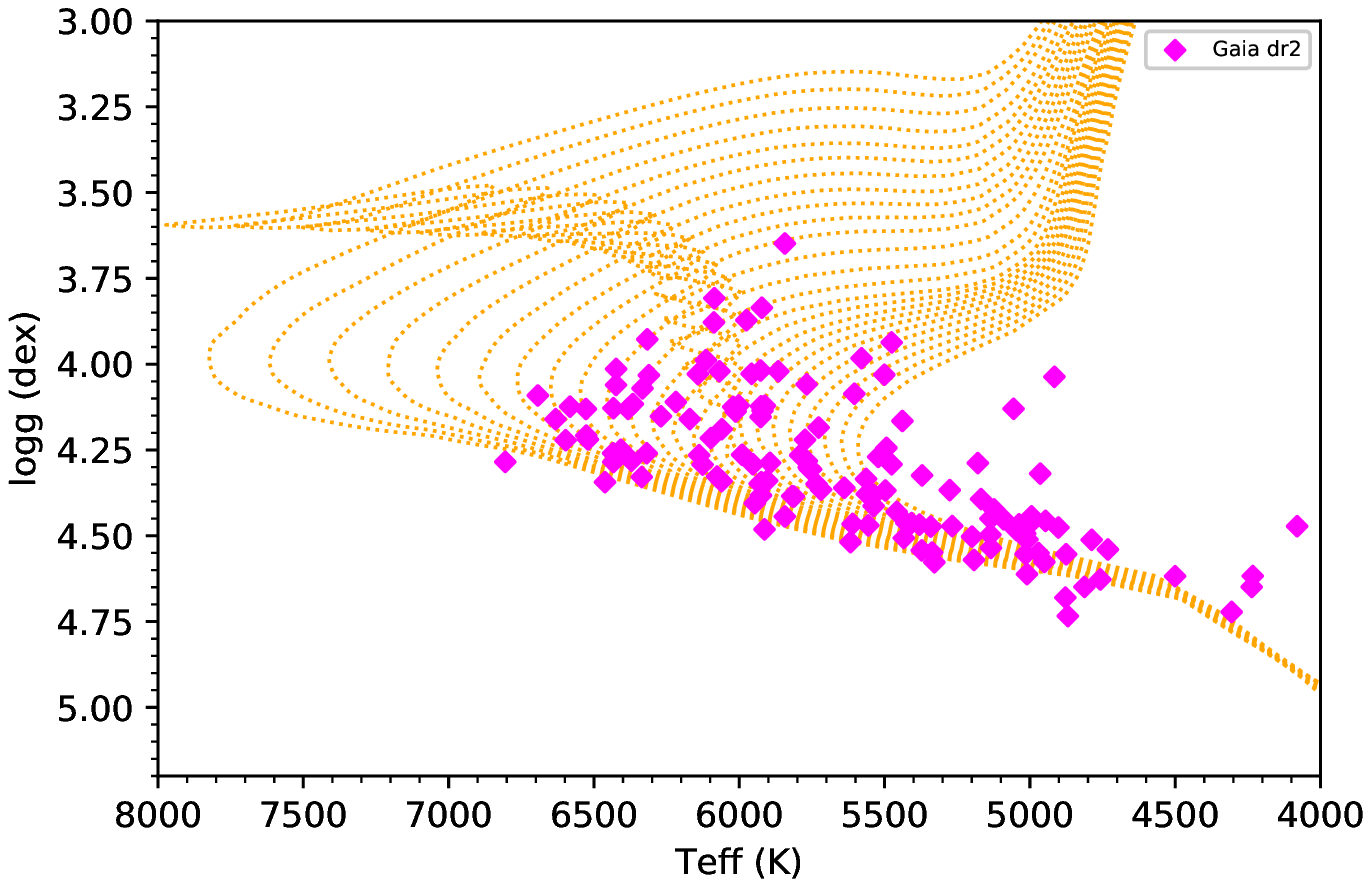}
\caption{Kiel diagram (log~g vs T$_{\rm eff}$) using values listed in SWEET-Cat, obtained by FAMA, FASMA and derived by {\it Gaia}. The outermost tracks correspond to log(age/yr)$=8.95$ (bottom track) and $10.15$ (upper track) with [M/H]$=0.058$, $Z=0.0198$, $Y=0.273$.}
\label{fig:6}       
\end{figure*}

 Another important check is based on the comparison with theoretical isochrones, computed in
T$_{\rm eff}$ vs log~g plane, for a set of ages, keeping the metallicity constant at [M/H]=0.058
\footnote{Most of our targets have indeed close-to-Solar metallicity, with a mean [Fe/H] of the sample +0.11$\pm$0.18~dex ([Fe/H] from Sweet-cat values)}. 
Fig.~\ref{fig:6} shows the Kiel diagrams (log~g vs T$_{\rm eff}$) for the values of log~g and T$_{\rm eff}$ listed in the Sweet-Cat, obtained by FAMA and by FASMA, and derived through {\it Gaia} photometry and parallax, respectively. 
The PARSEC isochrones \cite{Bressan2012} in a range of ages are over-plotted: 
the outermost tracks correspond to log(age/yr)$=8.95$ (bottom track) and $10.15$ (upper track) 
with [M/H]$=0.058$, $Z=0.0198$, $Y=0.273$.

Fig.~\ref{fig:6} (left top panel) points out how the SWEET-Cat gravities are not matching the isochrones path and appear to be overestimated at the highest temperatures and underestimated at the lowest temperatures.
In Fig.~\ref{fig:6} (right top panel) we present the same plot for the results obtained with FAMA: T$_{\rm eff}$ and log~g follow the expected trend, however there is an offset towards lower gravities.
In Fig.~\ref{fig:6} (left bottom panel) we show the results of FASMA. The dispersion is lower, however, at high temperature there is still an overestimation of log~g.
Finally, in Fig.~\ref{fig:6} (right bottom panel) the {\it Gaia} parameters are displayed, pointing out the best match with PARSEC isochrones.

\section{Conclusions}
\label{sec:5}
In our first test, we have compared the analysis of a sample of $\sim$150 spectra (high S/N and high spectral resolution).
In the parameters space close to Solar values the agreement among the three methods is good (T$_{\rm eff}$= $5000-6000$ K, log~g $=4.2-4.6$\,dex). 
However, at low and high temperatures some methods tend to under/overestimate the surface gravity log~g.
 This might have important effects on the derived stellar abundances.
External comparisons (using trigonometric log~g, log~g from light-curves and from isochrones) confirm the trends.
Corrections to these trends were already available (see, e.g, \cite{Mortier2014,Delgado2017}). 
We plan in the next tests to apply them,
using both corrections from asteroseismology and light-curves to provide more realistic gravites.
In this context we refer also to the wider discussion present in literature on the disagreement 
between spectroscopic and evolutionary or photometric log~g values (e.g. see \cite{Torres2012,Sozzetti2007}).

However, it is important to notice that, despite the differences in the derived photometric parameters, 
especially in log~g, the three spectroscopic methods agree very well on the final metallicity, except in the hottest temperature regime.
Due to high quality of the {\it Gaia} photometry and parallax for the Ariel targets, a viable and welcome possibility is to adopt the surface gravity
homogeneously derived from {\it Gaia} to compute chemical abundances.


%
%

\begin{acknowledgements}
We like to thank the anonymous referees for the fruitful comments on our paper.\\
AB, MT, LM, MvS, GC acknowledge  the funding from MIUR Premiale 2016: MITIC.\\ 
E.D.M., V.A. and S.G.S. acknowledge the support from Funda\c{c}\~ao para a Ci\^encia e a Tecnologia (FCT) through national funds and from FEDER through COMPETE2020 by the following grants: UID/FIS/04434/2019, UIDB/04434/2020\\ and UIDP/04434/2020;
PTDC/FIS-AST/32113/2017 \\and POCI-01-0145-FEDER-032113;\\ PTDC/FIS-AST/28953/2017 
\\and POCI-01-0145-FEDER-028953.\\ 
https://www.overleaf.com/project/5ee29f0469a39e00013326b1.\\
E.D.M., V.A., S.G.S. also acknowledge the support from FCT through Investigador FCT contracts\\
IF/00849/2015/CP1273/CT0003,\\ IF/00650/2015/CP1273/CT0001,\\ IF/00028/2014/CP1215/CT0002.
\end{acknowledgements}

%
%
\newpage

\bibliographystyle{spmpsci}      
\bibliography{ArielBib.bib}   

%
%

\end{document}